\documentclass[twocolumn,twoside]{article}

\usepackage{titling}
\usepackage{authblk}
\usepackage{abstract}
\usepackage{graphicx}
\usepackage{geometry}
\usepackage{url}
\usepackage{lipsum}
\usepackage{cite}
\usepackage{amsmath}
\usepackage{bm}

\pretitle{\begin{center}\LARGE\bfseries}
\posttitle{\par\end{center}}
\preauthor{\begin{center}\large}
\postauthor{\par\end{center}}
\predate{\begin{center}\large}
\postdate{\par\end{center}}

\title{Eight-shot measurement of spatially non-stationary complex coherence function}
\author[1,$\dagger$]{Pranay Mohta}
\author[1]{Abhinandan Bhattacharjee\thanks{Present address: Paderborn University, Integrated Quantum Optics, Institute for Photonic Quantum Systems, Warburger Straße 100, 33098 Paderborn, Germany.}}
\author[1,**]{Anand K. Jha}
\affil[1]{Department of Physics, Indian Institute of Technology Kanpur, Kanpur, Uttar Pradesh, 208016, India}

\affil[$\dagger$]{mohtapranay01@gmail.com}
\affil[**]{akjha@iitk.ac.in}

\usepackage{hyperref}
\hypersetup{
    colorlinks=true,
    linkcolor=blue,
    filecolor=magenta,      
    urlcolor=cyan,
    citecolor=blue
}

\begin{document}

\twocolumn[
  \maketitle
  \begin{abstract}
    Spatial coherence plays an important role in several real-world applications ranging from imaging to communication. As a result, its accurate characterization and measurement are extremely crucial for its optimal application. However, efficient measurement of an arbitrary complex spatial coherence function is still very challenging. In this letter, we propose an efficient, noise-insensitive interferometric technique that combines wavefront shearing and inversion for measuring the complex cross-spectral density function of the class of fields, in which the cross-spectral density function depends either on the difference of the spatial coordinates, or the squares of spatial coordinates, or both. This class of fields are most commonly encountered, and we  experimentally demonstrate high-fidelity measurement of many stationary and non-stationary fields.
  \end{abstract}
]

Characterizing the spatial coherence of optical fields is a fundamental pursuit in optics, with implications ranging from imaging to communication. The spatial coherence, quantified by the cross-spectral density function, is a key descriptor of correlations between two spatial locations in an optical field. Spatially partially coherent fields find various applications such as optical coherence tomography (OCT) \cite{dhalla2010ol}, imaging through turbulence \cite{redding2012np, redding2015pnas}, coherence holography \cite{naik2009oe}, photon correlation holography \cite{naik2011oe}, optical communication \cite{ricklin2002josaa}, and particle trapping \cite{dong2012pra}. 

While many spatially partially coherent light sources utilized in practical applications are spatially stationary, the spatially non-stationary sources are often encountered in real-world situations and are equally important for applications. For instance, the simplest optical system, a thin lens, introduces a quadratic phase in an optical field passing through it and is thus spatially non-stationary. The coherence function for free space propagation of a spatially incoherent source is of quadratic nature (see Section 5.7 of \cite{goodman2015statistical}). In non-line-of-sight imaging, the coherence function exhibits a quadratic nature, holding crucial information about the object's distance from the detector \cite{batarseh2018nc}. Recent research has seen a surge of interest in non-uniform coherence function, specifically those dependent on the difference of square of spatial coordinates, due to their unique propagation characteristics. These beams find applications in coherence-based optical encryption \cite{peng2021px}, simultaneous optical trapping of two types of Rayleigh particles with different refractive indices \cite{liu2015oc}, and the generation of self-focusing \cite{lajunen2011ol} and multi-focusing beams \cite{xu2023ol}.

Existing techniques for measuring the spatial coherence function such as the Young’s double-slit interferometer \cite{turunen1991josaa} and its variants \cite{santarsiero2006ol} suffer from drawbacks such as low light efficiency, long measurement time and stability requirements. Alternative methods, including shearing interferometer\cite{efimov2013ol}, sagnac interferometer\cite{rezvaniNaraghi2017ol}, phase-space tomography \cite{nugent1992prl}, shadows- \cite{wood2014ol}, gratings- \cite{koivurova2017ao}, masks \cite{mejia2007oc}, obstacles- \cite{hooshmandZiafi2020ol} and fiber optics-based \cite{anderson1993oengg} methods have been introduced. However, these methods often work well only for the coherence function of a specific form or reliably measure only its amplitude. Free-space propagation methods \cite{rydberg2007oe, petruccelli2013oe} require prior knowledge of the coherence properties and their iterative phase retrieval method may not converge to the correct form. Efficient coherence function measurement was demonstrated in \cite{bhattacharjee2018apl} but was limited to real and spatially stationary coherence functions. The work in \cite{halder2020ol} extended this idea to arbitrary coherence functions for both spatially stationary and non-stationary fields but required hundreds of measurements, compromising efficiency and system compactness. Recent study \cite{torcalMilla2023ol} addressed the compactness issue but still require a large number of measurements.  In contrast, in this letter, we report experimental demonstration of an efficient noise-insensitive technique for measuring the complex cross-spectral density functions of a broad class of cross-spectral density functions, which depend either on the difference of spatial coordinates, or squares of spatial coordinates, or both.

\begin{figure*}[!t]
\centering
\includegraphics[scale=0.33]{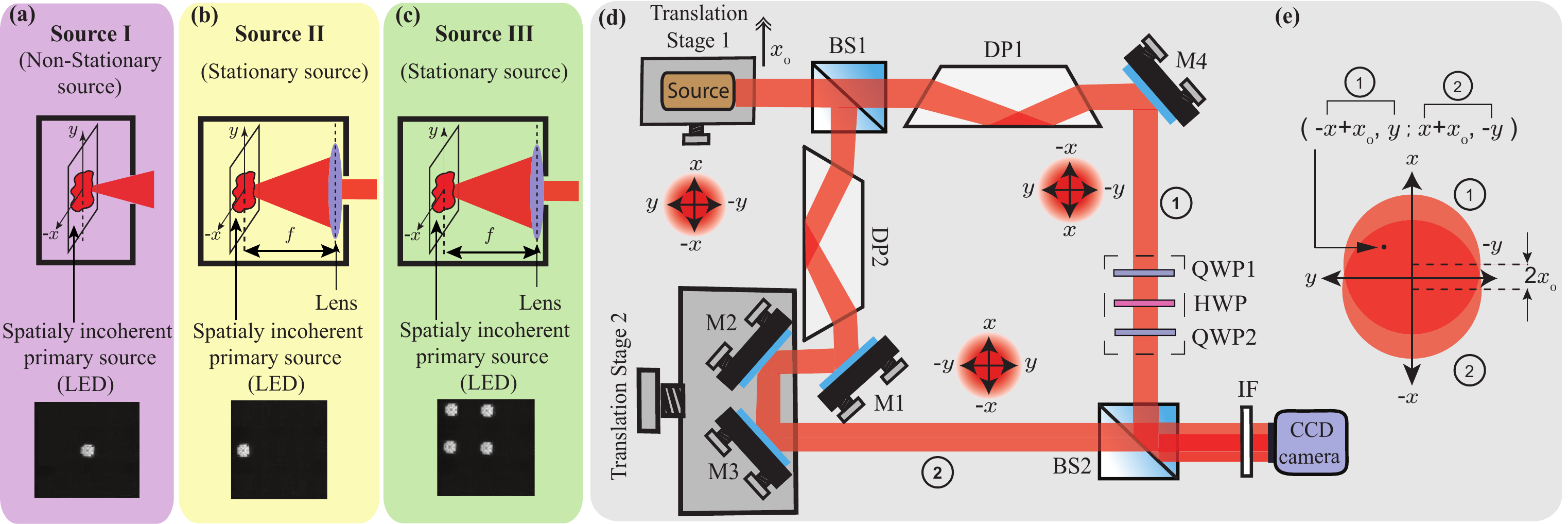}
\caption{(a), (b) and (c) are the three light sources. (d) Schematic diagram of the experimental setup. There is a dove prism (DP) in each arm, at $90^o$ to one another. Arm 1 contains a geometric phase unit consisting of QWP1, HWP, and HWP2. An interference filter (IF) centered at 632nm having a wavelength bandwidth of 10 nm is used before the CCD camera. (e) The two interfering wavefronts at the CCD camera plane. 
BS stands for beam splitter, M for mirror, QWP for quarter-wave plate, and HWP for half-wave plate.}
\label{fig1}
\end{figure*}

Fig.~\ref{fig1} shows the schematic diagram of our proposed experimental technique, which involves a modified Mach-Zehnder type interferometer. The cross-spectral density function of the field at two space points $\boldsymbol{\rho_1}$ and $\boldsymbol{\rho_2}$, is defined as $W(\boldsymbol{\rho_1},\boldsymbol{\rho_2}) = \langle E_{in}^*(\boldsymbol{\rho_1})E_{in}(\boldsymbol{\rho_2}) \rangle$, where $\langle \cdots\rangle$ denotes ensemble average over various realizations of the field, and $E_{in}(\boldsymbol{\rho_1})$, and $E_{in}(\boldsymbol{\rho_2})$ are the electric fields at the two spatial locations. One of the dove prisms inverts the incoming wavefront about the x-axis and the other one about the y-axis. As we work with linearly polarized light, the chosen configurations of the dove prisms does not alter the polarization state \cite{karan2022ao}. Arm-1 of the interferometer contains a geometric phase unit comprising two quarter-wave plates oriented at $45^o$ with respect to the polarization of light and a half-wave plate (HWP) in between. Rotating the HWP by an angle $\theta$ introduces a geometric phase of $2\theta$ \cite{jha2008prl}. Translating the source in the $x$-direction by $x_0$, we introduce a shear of $2x_0$ between the two interfering wavefronts. Fig.~\ref{fig1}(e) illustrates the resulting wavefronts at the output port of the interferometer, which can be expressed as $
E_{\rm out}(\boldsymbol{\rho}) \equiv E_{\rm out}(x,y)= k_1E_{\rm in}(-x+x_0,y)e^{i(\omega_0t_1+\beta_1)} 
 + k_2E_{\rm in}(x+x_0,-y)e^{i(\omega_0t_2+\beta_2)}$,
\begin{figure*}[ht]
\centering
\includegraphics[scale=0.32]{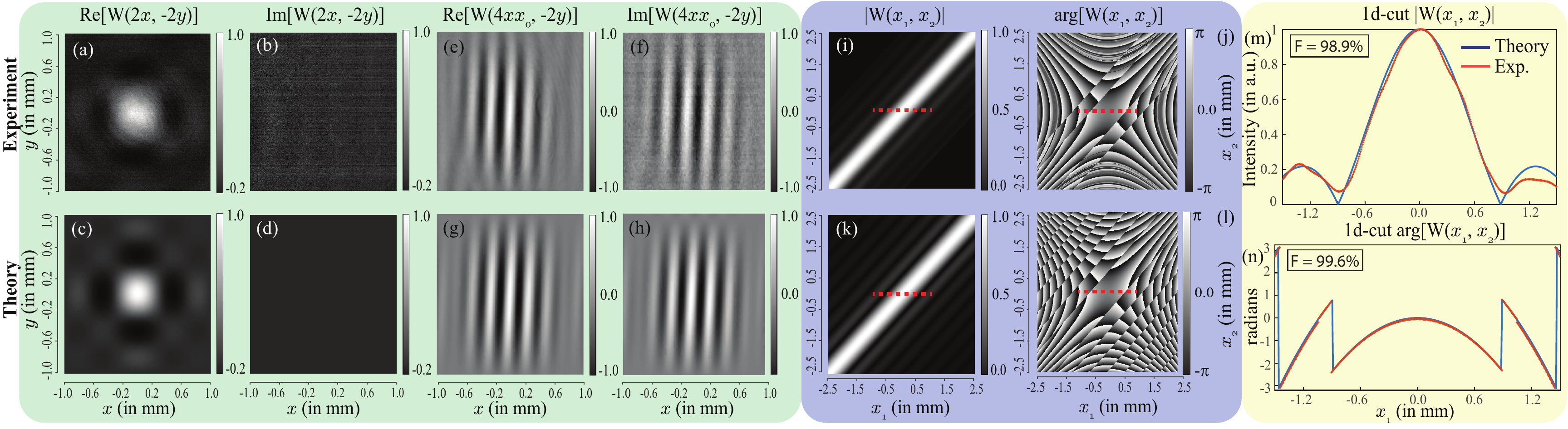}
\caption{Source I: (a) and (b) The experimentally measured $\text {Re}[W(-x+x_0,y;x+x_0,-y)]$ and $\text {Im}[W(-x+x_0,y;x+x_0,-y)]$ (c) and (d) The theoretical $\text {Re}[W(-x+x_0,y;x+x_0,-y)]$ and $\text {Im}[W(-x+x_0,y;x+x_0,-y)]$ with no shear, i.e., $x_0=0$ and $y_0=0$. (e)-(h) is the same with  shear $x_0=2.5\rm mm$ and $y_0=0\rm mm$.
(i) and (j) The experimentally reconstructed $|W(x_1,x_2)|$ and $\text {arg}[W(x_1,x_2)]$. (k) and (l) The theoretical $|W(x_1,x_2)|$ and $\text {arg}[W(x_1,x_2)]$.
(m) and (n) The 1d-cut of $|W(x_1,x_2)|$ and $\text {arg}[W(x_1,x_2)]$ along the dotted red line.}
\label{fig2}
\end{figure*}
\begin{figure*}[ht!]
\centering
\includegraphics[scale=0.40]{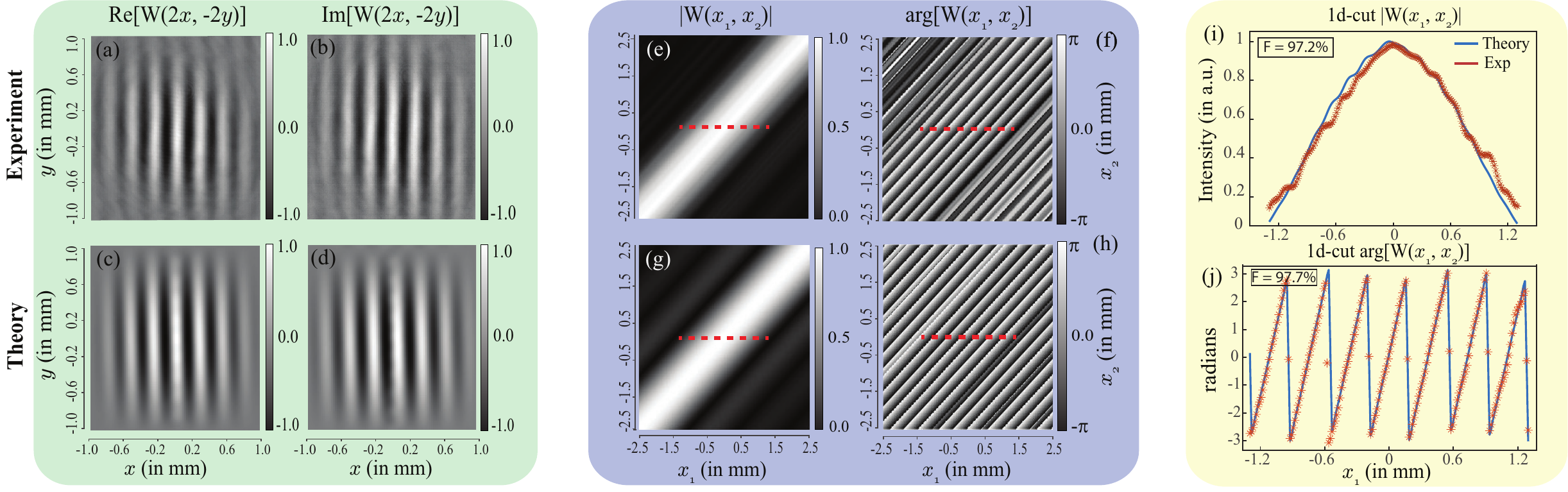}
\caption{ Source II: (a) and (b) The measured $\text {Re}[W(2x,-2y)]$ and $\text {Im}[W(2x,-2y)]$. (c) and (d) The theoretical $\text {Re}[W(2x,-2y)]$ and $\text {Im}[W(2x,-2y)]$.
(e) and (f) The experimentally reconstructed $|W(x_1,x_2)|$ and $\text {arg}[W(x_1,x_2)]$. (g) and (h) The theoretical $|W(x_1,x_2)|$ and $\text {arg}[W(x_1,x_2)]$.
(i) and (j) The 1d-cut of $|W(x_1,x_2)|$ and $\text {arg}[W(x_1,x_2)]$ along the dotted red line.}
\label{fig3}
\end{figure*}
\begin{figure*}[!t]
\centering
\includegraphics[scale=0.40]{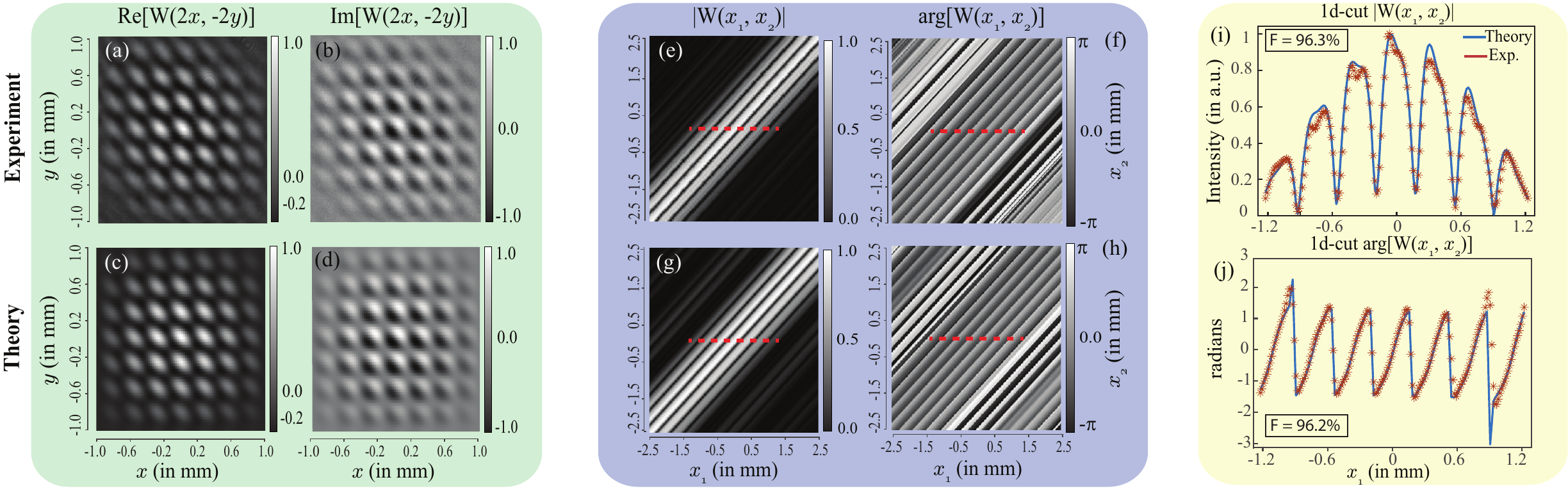}
\caption{ Source III: (a) and (b) The measured $\text {Re}[W(2x,-2y)]$ and $\text {Im}[W(2x,-2y)]$. (c) and (d) The theoretical $\text {Re}[W(2x,-2y)]$ and $\text {Im}[W(2x,-2y)]$.
(e) and (f) The experimentally reconstructed $|W(x_1,x_2)|$ and $\text {arg}[W(x_1,x_2)]$. (g) and (h) The theoretical $|W(x_1,x_2)|$ and $\text {arg}[W(x_1,x_2)]$.
(i) and (j) The 1d-cut of $|W(x_1,x_2)|$ and $\text {arg}[W(x_1,x_2)]$ along the dotted red line.}
\label{fig4}
\end{figure*}
where the time taken by the field to travel through the two arms is $t_1$ and $t_2$; $\omega_0$ is the cnetral frequency of the field; $\beta_1$ and $\beta_2$ are phases other than the dynamical phase acquired by the field; and $k_1$ and $k_2$ are the constants. We assume that $t_2-t_1$ is less than the coherence time of the field. The intensity $I_{\rm out}(x, y)\equiv \langle E^*_{\rm out}(x, y) E_{\rm out}(x, y) \rangle$ at the output of the interferometer is given by
\begin{multline}
I_{\rm out}(x, y) = |k_1|^2I_1(-x+x_0,y) + |k_2|^2I_2(x+x_0,-y)\\ +2k_1k_2Re[W(-x+x_0,y;x+x_0,-y)]\cos\delta \\ -2k_1k_2Im[W(-x+x_0,y;x+x_0,-y)]\sin\delta + I_{\rm b}(x, y), \label{eq2}
\end{multline}
where $\delta=\omega_{0}(t_2-t_1)+(\beta_2-\beta_1)$. We express $\langle E_{in}^*(-x+x_0,y)E_{in}(-x+x_0,y)\rangle=I_1(-x+x_0,y)$, $\langle E_{in}^*(x+x_0,-y)E_{in}(x+x_0,-y)\rangle=I_2(x+x_0,-y)$ and $W(-x+x_0,y;x+x_0,-y)=\langle E_{in}^*(-x+x_0,y)E_{in}(x+x_0,-y) \rangle$. Re[$\cdots$] and Im[$\cdots$] denote the real and imaginary parts. The contribution $I_{\rm b}^{\delta}(x, y)$ is the background noise. If $I^{\delta_c}_{\rm out}(x,y)$ and $I^{\delta_d}_{\rm out}(x, y)$ are the output intensities measured at $\delta=\delta_c$ and $\delta=\delta_d$, the difference intensity $\Delta I^{(\delta_c, \delta_d)}_{\rm out}(x, y)=I^{\delta_c}_{\rm out}(x, y)-I^{\delta_d}_{\rm out}(x, y)$ can be written as
\begin{multline}
\Delta I^{(\delta_c, \delta_d)}_{\rm out}(x, y)=  I_{\rm b}^{\delta_c}(x, y) - I_{\rm b}^{\delta_d}(x, y) \\ + 2k_1k_2 \lbrace Re[W(-x+x_0,y;x+x_0,-y)](\cos\delta_c-\cos\delta_d)  \\
-Im[W(-x+x_0,y;x+x_0,-y)](\sin\delta_c-\sin\delta_d)\rbrace.\label{eq3}
\end{multline}
Since the background noise does not depend on $\delta$, we get $I_{\rm b}^{\delta_c}(x, y) - I_{\rm b}^{\delta_d}(x, y)\approx 0$. Therefore, setting $\delta_c=0$  and $\delta_d=\pi$, we get Re[$W(-x+x_0,y;x+x_0,-y)$]$= \Delta I_{\rm out}^{(0,\pi)}(x, y) / 4k_1k_2$; and setting $\delta_c=3\pi/2$ and $\delta_d=\pi/2$, we get Im[$W(-x+x_0,y;x+x_0,-y)$]$= \Delta I^{(3\pi/2, \pi/2)}_{\rm out}(x, y) / 4k_1k_2$. Thus, by measuring the difference intensities $\Delta I_{\rm out}^{(0,\pi)}(x, y)$ and $\Delta I^{(3\pi/2, \pi/2)}_{\rm out}(x, y)$, we obtain the real and imaginary parts to within the same scaling factor $4k_1k_2$. This way, we obtain the complex cross-spectral density function $W(-x+x_0,y;x+x_0,-y)$ in a four-shot manner, which not only obviates the necessity for the precise knowledge of $k_1$, $k_2$, $I_1$ and $I_2$ but also becomes completely insensitive to background noise \cite{bhattacharjee2018apl, kulkarni2017nc, kulkarni2018pra, karan2023prapplied}. Knowing the real and imaginary parts, one can find the amplitude and phase as: $|W|=\sqrt{(\Delta I_{\rm out}^{(0,\pi)})^2 + (\Delta I_{\rm out}^{(3\pi/2,\pi/2)})^2}/(4k_1k_2)$, and ${\rm arg}[W]=\tan^{-1}\left(\Delta I_{\rm out}^{(3\pi/2,\pi/2)}/\Delta I_{\rm out}^{(0,\pi)}\right)$.

Our aim is to measure a class of spatially non-stationary field for which the complex cross-spectral density function factorizes as $W(x_1, x_2; y_1, y_2)=W_x(x_1, x_2)W_y(y_1, y_2)$, with the cross-spectral density in $x$- and $y$-directions being of the form $W(x_1, x_2)=W_1(x_2-x_1)W_2(x^2_2-x^2_1)$. This is the most commonly-encountered class of fields \cite{batarseh2018nc, peng2021px, liu2015oc, lajunen2011ol, xu2023ol}. It becomes spatially stationary in situations in which $W_2(x^2_2-x^2_1)=1$. In what follows, for conceptual clarity, we only consider the $x$-coordinate, noting that the same principles can be extended to the $y$-coordinate. Setting $x_2=x+x_0$ and $x_1=-x+x_0$ with the shear $x_0=0$ and thus taking $W_2(0)=1$, we obtain
\begin{multline}
W(-x, x)=W_1(2x)W_2(0)=W_1(2x) = A_1(2x)e^{i\phi_1(2x)}.
\label{eq4}
\end{multline}
By employing the above four-shot technique, one can measure $W(-x, x)=W_1(2x)$ and thereby the amplitude $A_1(2x)$ and phase $\phi_1(2x)$. The phase $\phi_1(2x)$, which is indeterminate up to $2\pi$, is unwrapped and we obtain the phase $\phi^u_1$, which ranges from $-\infty$ to $\infty$. The function $W_1(x_2-x_1)$ is reconstructed by mapping $W_1(2x)$ to $W_1(x_2-x_1)$ with $ x=(x_2-x_1)/2$. Next, setting $x_2=x+x_0$ and $x_1=-x+x_0$, with $x_0\neq0$, we obtain
\begin{multline}
W(-x+x_0, x+x_0)=W_1(2x)W_2(4xx_0)\\ =A_1(2x)A_2(4xx_0)e^{i[\phi_1(2x)+\phi_2(4xx_0)]}
\label{eq5}
\end{multline}
Again, by employing the above four-shot technique, one can measure $W(-x+x_0, x+x_0)$ and thereby the amplitude $A_2(4xx_0)$ as $A_2(4xx_0)=|W(-x+x_0, x+x_0)|/|W(-x, x)|$. We note that to deal with the small values of $|W(-x, x)|$ in the denominator, we set an appropriate threshold or floor value below which all the values are set to the floor value. We define $\phi^u$ as the unwrapped version of $\phi_1(2x)+\phi_2(4xx_0)$. The unwrapped phase $\phi^u_2=\phi^u-\phi^u_1$ then yields $\phi_2(4xx_0)$. The function $W_2(x^2_2-x^2_1)$ is reconstructed by mapping $W_2(4xx_0)$ to $W_2(x^2_2-x^2_1)$ with $x=(x^2_2-x^2_1)/4x_0$. This way the  function $W(x_1,x_2)$ is reconstructed with 8-shot intensity measurement.  We note that the shear size $x_0$ should be smaller than the spatial coherence length of the field for a better interference visibility, and it should be large enough for the imaging system to be able to spatially resolve it. Other than this, the measurements are independent of $x_0$. We further note that, although for measuring a spatially non-stationary $W(x_1, x_2)$ one requires 8-shot intensity measurements, a spatially stationary cross-spectral density function can be measured using only four intensity shots.

In our experiments, we use a commercially available planar light-emitting diode (LED) unit consisting of 9 separate LEDs arranged in a $3\times 3$ square grid. Each LED has dimension $0.8$ mm $\times$ $0.8$ mm, with a separation of 1.9 mm (see Fig.~{\ref{fig1}}). The phase $\delta$ is changed using the geometric phase unit in arm-1 so that $t_2-t_1$ and thereby the degree of temporal coherence remains unaffected. To find $\delta=0$, we send a known field through the interferometer and adjust the geometric phase such that the output intensity at the camera is maximum. Once $\delta=0$ configuration is known, we switch on the field of interest and obtain different $\delta$ values by rotating the HWP. Source I is just a single LED unit as shown in Fig.~\ref{fig1} (a). As LED is a spatially incoherent source, its cross-spectral density is calculated using the van Cittert-Zernike theorem and is of the form $W(x_1,y_1;x_2,y_2)=W_1(x_2-x_1,y_2-y_1)W_2(x^2_2-x^2_1,y^2_2-y^2_1)$ \cite{goodman2015statistical}. Two four-shot measurements are taken, one with $x_0=0; y_0=0$ and another with shear $x_0=2.5$ mm; $y_0=0$ mm. Shear is introduced using the translations stage-1 solely along the $x$-axis. However, simultaneous shear along both the $x$ and $y$-directions can be achieved using an xy-translation stage and thus the cross-spectral density as a function of both  $x$ and $y$ can be measured. Source II shown in Figs.~\ref{fig1}(b) and source III shown in Figs.~\ref{fig1}(c) consist of one off-axis LED and a group of four off-axis LEDs, respectively, kept at the back focal plane of a plano-convex lens of focal length $45$ cm. Both sources are spatially stationary and have the cross spectral density function of the form $W(x_1,x_2)=W(x_2-x_1)$ and is calculated by taking the Fourier transform of the spectral density of the LED \cite{aarav2017pra, bhattacharjee2018apl}.

The experimental results and analysis corresponding to source I, source II and source III have been presented in Fig.~\ref{fig2}, Fig.~\ref{fig3}, and \ref{fig4}, respectively. In Figs.~\ref{fig2}-\ref{fig4}, the theory have been obtained by taking the actual image of the three LED sources and then calculating the far-field cross-spectral density function using the formulation given in Refs.~\cite{aarav2017pra, bhattacharjee2018apl} for source I and in Ref.~\cite{goodman2015statistical} for Source II and III. (see Supplemental document section 2). In Figs.~\ref{fig2}-\ref{fig4}, both theory and experimental plots for the real and imaginary parts as well as for the amplitude of the cross spectral density function have been scaled such that the maximum value of the plot is equal 1. For plotting the phase, no scaling is done. The fidelity of our experimental results, calculated using the $R^2$ coefficient ~\cite{karan2023prapplied} and indicated in Figs.(~\ref{fig2}-\ref{fig4}) are greater than $96\%$, showing excellent match between theory and experiment. We note that the accurate experimental realization of the phase $\delta$ through our geometric phase setup is very important for the accurate reconstruction of the coherence function. Nonetheless, we find that an error of $\theta$ in the geometric phase has no effect on the reconstructed amplitude of the cross-spectral density function while the error in the reconstructed phase increases only linearly with $\theta$ (see Supplemental document section 1).

In summary, we have proposed and demonstrated an  efficient high-fidelity technique based on wavefront-inversion and shearing for measuring the complex cross-spectral density function of both stationary and non-stationary spatially partially coherent optical fields in just eight shots. The technique is insensitive to background noise, takes only about eight seconds for one complete dataset and has the potential to serve as a valuable tool for understanding and leveraging coherence characteristics in diverse practical applications. 

\section*{Supplemental document}
\section{Error analysis for geometric phase}
From Eq.~ref{2} in the main text we see that the difference in the intensities of the two interferograms is,
\begin{multline}
\Delta I_\text{out}(x, y)=I^{\delta_c}_\text{out}(x, y)-I^{\delta_d}_\text{out}(x, y)
\\=2k_1k_2 \lbrace \text{Re}[\text{W}(-x+x_0,y;x+x_0,-y)](\text{cos}(\delta_c)-\text{cos}(\delta_d)) 
\\-\text{Im}[\text{W}(-x+x_0,y;x+x_0,-y)](\text{sin}(\delta_c)-\text{sin}(\delta_d))\rbrace.\label{eq1s}
\end{multline}
The geometric phase setup, depicted in arm 1 of Figure 1(d) in the main text, involves two quarter-waveplates (QWP1) oriented at 45° with respect to the polarization of light, with a half-waveplate (HWP) placed in between. Rotating the HWP by an angle $\phi$ introduces a relative phase difference of $2\phi$ between the two arms, denoted as $\delta = 2\phi $. To find $\delta=0$, we send a known field through the interferometer and adjust the geometric phase such that the output intensity at the camera is maximum. Once $\delta=0$ configuration is known, we switch on the field of interest and obtain different $\delta$ values by rotating the HWP. The precision of $\delta$ relies on the accuracy of the HWP rotation. In our setup, the HWP was affixed to an automated rotation stage, ensuring uniform error across all measurements.
Let $\theta$ be the error in setting the phase difference between the two arms, $\delta_{c,d}$. Then, 
\begin{multline}
\Delta I_\text{out}(x, y)=2k_1k_2[W_\text{R}\lbrace \text{cos} \delta_c \text{cos} \theta-\text{sin} \delta_c \text{sin} \theta - \text{cos} \delta_d \text{cos} \theta+\\ \text{sin} \delta_d \text{sin} \theta \rbrace 
\\-W_\text{Im}\lbrace \text{sin} \delta_c \text{cos} \theta + \text{cos} \delta_c \text{sin} \theta - \text{sin} \delta_d \text{cos} \theta \\-\text{cos} \delta_d \text{sin} \theta \rbrace].\label{eq2s}
\end{multline}
where $W_\text{R} =\text{Re}[W(-x+x_0,y;x+x_0,-y)] $ and $W_\text{Im}=\text{Im}[W(-x+x_0,y;x+x_0,-y)]$. From here onwards we adopt this renaming for brevity.

Putting $\delta_c=0$ and $\delta_d=\pi$ we get,
\begin{equation}
\Delta I_\text{out}(x, y)|_1=4k_1k_2 \lbrace (\text{cos} \theta)W_\text{R}-(\text{sin} \theta)W_\text{Im}\rbrace.\label{eq3s}
\end{equation}
Putting $\delta_c=3\pi/2$ and $\delta_d=\pi/2$ we get,
\begin{equation}
\Delta I_\text{out}(x, y)|_2=4k_1k_2 \lbrace (\text{sin} \theta)W_\text{R}+(\text{cos} \theta)W_\text{Im}\rbrace.\label{eq4s}
\end{equation}
Squaring and adding Eqs. (\ref{eq3s}) and (\ref{eq4s}) we get,
\begin{equation}\label{eq5s}
\begin{split}
\Delta I^2_\text{out}(x, y)|_1+\Delta I^2_{out}(x, y)|_2&=(4k_1k_2 )^2 (W_\text{R}^2
+W_\text{Im}^2)
\\&=16k^2_1k^2_2|W|^2 
\end{split}
\end{equation}
where $|W|$ is the amplitude of the cross-spectral density function. So we can write,
\begin{equation}
|W|=\frac{\sqrt{\Delta I^2_\text{out}(x, y)|_1+\Delta I^2_{out}(x, y)|_2}}{4k_1k_2} \label{eq6}
\end{equation}
We see that amplitude of the cross-spectral density function is unaffected by the error in $\delta$.

From Eqs. (\ref{eq3s}) and (\ref{eq4s}) we see that the phase profile of the cross-spectral function is given by,
\begin{equation}
\text{arg[W]}=\text{tan}^{-1}\left[\frac{\Delta I_\text{out}(\boldsymbol{\rho})|_2}{\Delta I_\text{out}(\boldsymbol{\rho})|_1}\right]=\text{tan}^{-1}\left[\frac{W_\text{Im}+(\text{tan}\theta) W_\text{R}}{W_\text{R}-(\text{tan}\theta) W_\text{Im}}\right]\label{eq7}
\end{equation}
Therefore the error in the phase profile of the cross-spectral density function is,
\begin{equation}
\text{Error(arg[W])}=\text{tan}^{-1}\left[\frac{W_\text{Im}+(\text{tan}\theta) W_\text{R}}{W_\text{R}-(\text{tan}\theta) W_\text{Im}}\right] - \text{tan}^{-1}\left[\frac{W_\text{Im}}{W_\text{R}}\right] \\=\theta\label{eq8}
\end{equation}
We find that an error of $\theta$ in the geometric phase has no effect on the reconstructed amplitude of the cross-spectral density function while the error in the reconstructed phase increases only linearly with $\theta$.
\section{Cross-spectral density of employed light sources}
\subsection{Source I}
The subsequent analysis is adapted from section 5.7 of Ref.~\cite{goodman2015statistical}.
For Source I we use just a LED which can reasonably be modelled as an extended collection of independent radiators. So it can be considered as a spatially incoherent source. The Van Cittert-Zernike theorem answers how exactly the cross-spectral density propagates away from the source.
For a quasi-monochromatic light, the spatial cross-spectral density function propagates according to the law,
\begin{multline}
W(Q_1,Q_2) = \iiiint_\Sigma W(P_1,P_2) \text{exp}\left[-\iota \frac{2\pi}{\lambda}(r_2-r_1) \right]
\\ \frac{\chi(\theta_1)}{\lambda r_1}\frac{\chi(\theta_2)}{\lambda r_2}dS_1 dS_2 \label{eq9}
\end{multline}
where $W(Q_1,Q_2)$ and $W(P_1,P_2)$ are the cross-spectral densities at the source and detection plane respectively. $\chi(\theta)$ is the obliquity factor where $\theta$ is the angle between the line joining $Q_1$ and $P_1$ and the normal to the surface $\Sigma$ at $P_1$.
\begin{figure}[t!]
\includegraphics{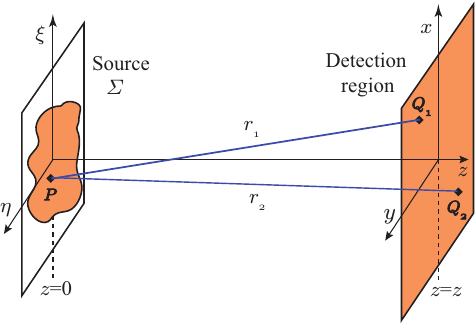}
\caption{Geometry for derivation of Van Cittert-Zernike theorem.}
\label{fig2s}
\end{figure}
For a spatially incoherent source, we have
\begin{equation}
W(P_1,P_2)=\kappa I(P_1)\delta(|P_1-P_2|)\label{eq10}
\end{equation}
Substituting this in Eq.(\ref{eq9}) gives,
\begin{multline}
W(Q_1,Q_2) =\frac{\kappa}{\lambda^2} \iint_\Sigma I(P_1) \text{exp}\left[-\iota \frac{2\pi (r_2-r_1)}{\lambda} \right]
\\ \frac{\chi(\theta_1)}{r_1}\frac{\chi(\theta_2)}{r_2}dS \label{eq11}
\end{multline}
where the geometry of the setting is illustrated in Fig.~\ref{fig2s}.
To simplify our expression, we employ the following assumptions and approximations:

1. Far-Field Approximation - The dimensions of both the source and detection area are significantly smaller than the separation distance $z$ between them.
\begin{equation}
\frac{1}{r_1 r_2} \approx \frac{1}{z^2}
\end{equation}
2. Small angle approximation, 
\begin{equation}
\chi(\theta_1) \approx \chi(\theta_1)\approx 1
\end{equation}
Following this, the cross-spectral density in the detection region takes the form
\begin{multline}
W(Q_1,Q_2) =\frac{\kappa}{(\lambda z)^2} \iint_\Sigma I(P_1) \text{exp}\left[-\iota \frac{2\pi (r_2-r_1)}{\lambda} \right]dS \label{14}
\end{multline}
In our further analysis, we make the assumption that the source and detection planes are parallel to each other, separated by a distance $z$. Additionally, we also employ the paraxial approximation. So,
\begin{multline}
r_1=\sqrt{z^2+(x_1-\xi)^2+(y_1-\eta)^2}\approx z+\frac{(x_1-\xi)^2+(y_1-\eta)^2}{2z}
\\r_2=\sqrt{z^2+(x_2-\xi)^2+(y_2-\eta)^2}\approx z+\frac{(x_2-\xi)^2+(y_2-\eta)^2}{2z}
\end{multline} 
Finally, we introduce the definitions $\Delta x=x_2-x_1$ and $\Delta y=y_2-y_1$, alongside the convention that $I(\xi,\eta)$ is zero for points $(\xi,\eta)$ outside the finite source region $\Sigma$. This leads us to the final expression of the Van Cittert-Zernike theorem:
\begin{multline}
W(x_1,y_1;x_2,y_2)=\frac{\kappa e^{\iota \psi}}{(\lambda z)^2} \iint_{-\infty}^{\infty} I(\xi,\eta) \\ \text{exp}\left[ \iota \frac{2\pi}{\lambda z}(\Delta x \xi+\Delta y \eta)\right]d\xi d\eta \label{eq16}
\end{multline}
where the phase factor $\psi$ is given by
\begin{equation}
\psi=\frac{\pi}{\lambda z}[(x^2_2+y^2_2)-(x^2_1+y^2_1)] \label{eq17}
\end{equation}
From Eq.(\ref{eq16}) and (\ref{eq17}) it is clear that the cross-spectral density of LED source is of the form $W(x_1,y_1;x_2,y_2)=W_1(x_2-x_1,y_2-y_1)W_2(x^2_2-x^2_1,y^2_2-y^2_1)$. 
\begin{figure}[b]
\includegraphics{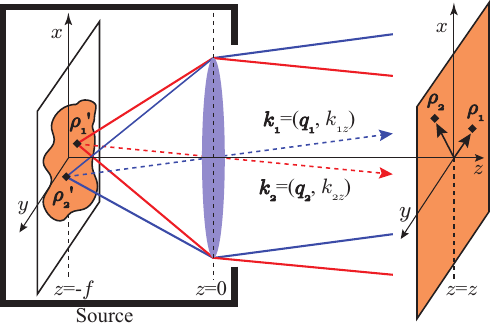}
\caption{Schematic illustration of generation of propagation-invariant spatially-stationary field using a spatially completely-uncorrelated primary source.}
\label{fig1s}
\end{figure}
\subsection{Source II \& III}The discussion that follows is sourced from Ref.~\cite{aarav2017pra}.
For Sources II and III, a monochromatic, planar, spatially completely incoherent primary source is positioned at the back focal plane, $z=-f$, of a lens located at $z=0$. Our spatially partially coherent field source consists of this composite system of the planar primary source and the lens. The field emanating from a spatial location $\bm\rho'$ at a distance $z$ is denoted as $V_s(\bm\rho', z)$. Given the complete spatial incoherence of our primary source, the fields $V_s(\bm\rho'_1, -f)$ and $V_s(\bm\rho'_2, -f)$ emerging from $\bm\rho'_1$ and $\bm\rho'_2$ at $z=-f$, respectively, are entirely uncorrelated. This lack of correlation is expressed as,
\begin{align}
\langle V_s^*(\bm\rho'_1, -f)V_s(\bm\rho'_2, -f)\rangle_e=I_s(\bm\rho'_1, -f)\delta(\bm\rho'_1-\bm\rho'_2).\label{position-correlation}
\end{align}
Here $I_s(\bm\rho'_1, -f)$ is the intensity of the primary source at $z=-f$. In our experiments, we use LEDs as our primary incoherent sources. Here, $I_s(\bm\rho'_1, -f)$ denotes the intensity of the primary source at $z=-f$. LED serve as our primary incoherent source as their positional correlations are modelled by the form outlined in Eq.~(\ref{position-correlation}) \cite{tziraki2000apb, pires2010optlett}.

For our primary source characterized by Eq.~(\ref{position-correlation}), each point on the source behaves as an independent point source. As each of these points is positioned at the back focal plane of a converging lens, the field $V_s(\bm\rho'_1, -f)$ originating from $\bm\rho'_1$ undergoes transformation into a plane wave with amplitude $a(\bm{q_1})$, where $\bm{q_1}$ denotes the transverse wave-vector associated with the plane wave \cite{born1999, goodman2005}. The lens effectively converts the non-correlation of the planar source in the position basis into non-correlation in the transverse wave-vector basis. The correlations between distinct transverse wave-vectors are quantified through the angular correlation function $\mathcal{A}(\bm{q_1}, \bm{q_2})$, defined as $\mathcal{A}(\bm{q_1}, \bm{q_2}) \equiv \langle a^*(\bm{q_1})a(\bm{q_2})\rangle_e$, where $\langle\cdots\rangle_e$ denotes the ensemble average. The angular correlation function for such a partially coherent source corresponds to the angular correlation function $\mathcal{A}(\bm{q_1}, \bm{q_2})$ at the exit face of the lens, i.e., at $z=0$, and is expressed as, 
\begin{align}
\mathcal{A}(\bm{q_1}, \bm{q_2}) \equiv \langle
a^*(\bm{q_1})a(\bm{q_2})\rangle_e=I_s(\bm{q_1})\delta(\bm{q_1}-\bm{q_2}).\label{angular-correlation-fn}
\end{align}   
Here, $I_s(\bm{q_1})$ represents the spectral density of the field, exhibiting the same functional form as the source intensity. As demonstrated below, this specific form of the angular correlation function is a prerequisite for the spatially-stationary and propagation-invariant characteristics of the partially-coherent field emerging from a source.

We proceed to derive the cross-spectral density function at $z=z$ generated by such a source. As detailed in Section 5.6 of Ref.~\cite{mandel}, assuming the plane-wave amplitude at $z=0$ is $a(\bm{q_1})$, the field $V(\bm\rho_1, z)$ at $z=z$ within the paraxial approximation is given by,
\begin{align}
V(\bm\rho_1, z)=e^{ik_0z}\iint_{-\infty}^{\infty}
a(\bm{q_1})e^{i\bm{q_1}.\bm\rho_1}e^{-i\tfrac{q_1^2 z}{2k_0}}d\bm{q_1}.
\end{align}
Here, $\bm{r_1}\equiv(\bm{\rho_1}, z)$,
$\bm{k_1}\equiv(\bm{q_1}, k_{1z})$, and $k_{1z}\approx k_1-q_1^2/(2 k_1)$ with $q_1=|\bm{q_1}|$ and $k_{1}=|\bm{k_1}|=k_0=\omega_0/c$, where $\omega_0$ is the frequency of the field. So the cross-spectral density function $W(\bm\rho_1, \bm\rho_2, z)\equiv \langle V^*(\bm\rho_1, z) V(\bm\rho_2, z) \rangle_e$ at $z=z$ is 
\begin{align}
W({\bm{\rho}_1}, {\bm{\rho}_2}, z)= &
\iint_{-\infty}^{\infty} \mathcal{A}(\bm{q_1}, \bm{q_2})
\notag\\ \times & e^{-i\bm{q_1}.\bm\rho_1+i\bm{q_2}.\bm\rho_2}
e^{-i\tfrac{(q_1^2-q_2^2)z}{2k_0}}d\bm{q_1} d\bm{q_2}.\label{cs-density}
\end{align}
Equation (\ref{cs-density}) dictates the evolution of spatial correlations in the field, as denoted by the cross-spectral density function, during propagation in the region $z>0$ after passing through the lens. Substituting the expression for the angular correlation function from Eq.~(\ref{angular-correlation-fn}) into Eq.~(\ref{cs-density}), we acquire,
\begin{align}
W({\bm{\rho}_1}, {\bm{\rho}_2}, z)= W(\bm\Delta\rho, z)= 
\int_{-\infty}^{\infty} I_s(\bm{q})
 e^{-i\bm{q.\Delta\rho}}d\bm{q},\label{cs-density2}
\end{align}
where $\bm\Delta\rho=\bm{\rho_1}-\bm{\rho_2}$. We see that the cross-spectral density function \(W(\bm\Delta\rho, z)\) in Eq.~(\ref{cs-density2}) follows coherent mode representation, where the coherent modes correspond to plane waves. The generated field exhibits the following characteristics: (1) \textit{Propagation invariance}---The cross-spectral density function and intensity remain constant regardless of changes in \(z\). (2) \textit{Spatial stationarity at a given \(z\), at least in a broad sense}---This is evident from the fact that the intensity \(I(\bm\rho, z)\) does not depend on \(\bm\rho\), while the cross-spectral density function depends only on \(\Delta\bm\rho\). (3) \textit{The cross-spectral density function \(W(\bm\Delta\rho, z)\) of the field is the Fourier transform of its spectral density \(I_s(\bm{q})\)}.  
Furthermore, as the spectral density mirrors the intensity pattern of the primary source, the cross-spectral density function of the field seamlessly transforms into the Fourier representation of the primary source's intensity profile. It's noteworthy that this methodology liberates the intensity function \(I_s(\bm{q_1})\) of the primary incoherent source from constraints — whether continuous, of finite dimensions, or structured as an ensemble of discrete points.

\section*{Funding}
Science and Engineering Research Board (Grant No. STR/2021/000035) and Department of Science \& Technology, Government of India (Grant No. DST/ICPS/QuST/Theme-1/2019).

\section*{Acknowledgements}
PM acknowledges the Prime Minister Research Fellowship (PMRF), Government of India.

\section*{Disclosures}
The authors declare no conflicts of interest.

\section*{Data Availability}
Data underlying the results presented in this paper are not publicly available at this time but may be obtained from the authors upon reasonable request.

\bibliographystyle{plain} 
\bibliography{8shot_refArxiv} 


\end{document}